\documentclass[preprint,preprintnumbers,amsmath,amssymb,show keys,showpacs,secnumarabic]{revtex4}

\usepackage[dvips]{graphicx}
\usepackage{amsmath,amsthm,amssymb}
\usepackage[dvips]{color}

\begin{document}

\title{The Grounds For Time Dependent Market Potentials \\From Dealers' Dynamics}%

\author{Kenta Yamada$^1$}\email[E-mail: ]{yamada@smp.dis.titech.ac.jp}
\author{Hideki Takayasu$^2$}
\author{Misako Takayasu$^1$}
\affiliation{$^1$Department of Computational Intelligence and Systems Science, Interdisciplinary Graduate School of Science and Engineering, Tokyo Institute of Technology, 4259 Nagatsuta-cho, Midori-ku, Yokohama 226-8502, Japan}

\affiliation{$^2$Sony Computer Science Laboratories, 3-14-13 Higashi-Gotanda, Shinagawa-ku, Tokyo 141-0022, Japan}

\begin{abstract}
We apply the potential force estimation method to artificial time series of market price produced by a deterministic dealer model. We find that dealers' feedback of linear prediction of market price based on the latest mean price changes plays the central role in the market's potential force. When markets are dominated by dealers with positive feedback the resulting potential force is repulsive, while the effect of negative feedback enhances the attractive potential force. 

\end{abstract}

\maketitle
\section{Introduction}
Improvement of time resolution of market data owing to the development of computer technology has been led to many new discoveries in econophysics. Empirical laws have been established in foreign exchange markets and stock markets such as the power laws in price change distributions, the long-tails in volatility auto-correlation, non-Poissonian transaction interval distributions, and so on \cite{empirical-laws1}\cite{volatility}\cite{transaction interval}. One of the goals of econophysics study may be to clarify the origin of these empirical laws. As the market prices are actually determined by the traders' transactions, it is natural to expect the origin of these empirical laws to be the dealers' actions. In our previous study of deterministic agent model which we call the dealer model, we successfully reproduced most of the basic empirical laws \cite{dealermodel0}\cite{dealermodel1}\cite{dealermodel2}. 

Recently, M.Takayasu and her coworkers introduced a new time series analysis method to observe non-trivial time dependent potential forces in market prices\cite{potentialmodel1}. It is shown that this potential force can be observed universally in financial markets, and the most of basic empirical laws are also satisfied in this formulation. In this paper we numerically observe potential forces of the dealer model and clarify an origin of market potentials.

\section{Potential forces observed in foreign exchange markets}\label{sec:RM-potential}
\begin{figure}[htbp]
  \begin{center}
    \includegraphics[width=13.0cm]{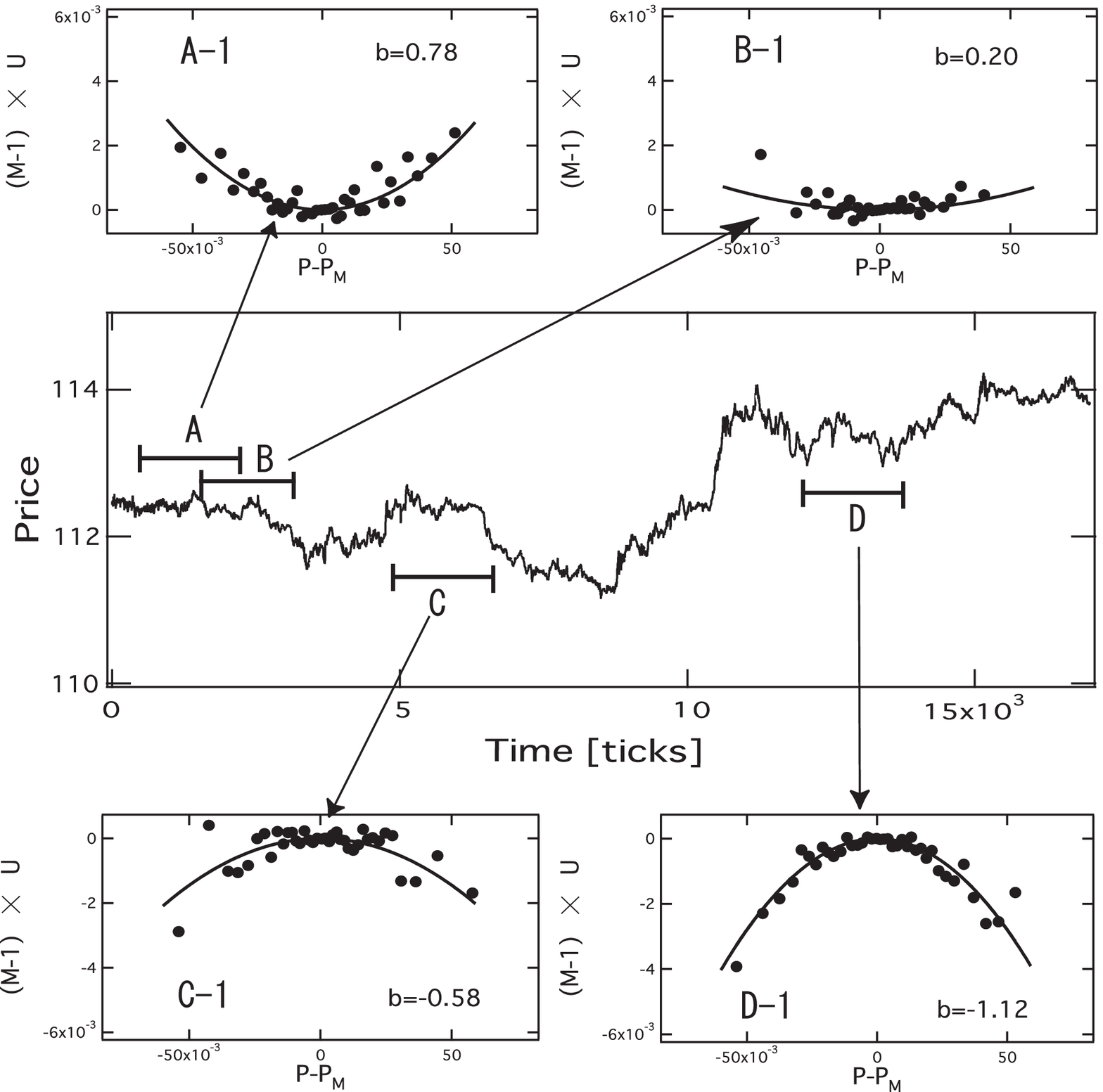}
  \end{center}
  \caption{Calculated potentials by Takayasu's method in a real market.The centered time series is noise free prices so-called optimal moving average during about two days from January 12th 1999 to January 14th 1999. The ranges marked A, B, C, D are all 2000 ticks and we calculated a potential for each range shown in A-1, B-1, C-1, D-1. In these potential figures, we set scales of horizontal axis and bottom axis the same size. We show the estimated potential curvature $b$ for each potential. }
  \label{fig:potential_bloom1.eps}
\end{figure}

In this section we briefly show an example of results of the potential analysis to a foreign exchange market data for review of the method and for comparison with the dealer model in the next section (Fig.\ref{fig:potential_bloom1.eps}). The figure located in the center is the optimal moving average price $P(u)$ defined by the tick time $u$ updated by every transaction. The optimal moving average is obtained by removing uncorrelated noise component\cite{OMA} from dollar-yen bid price during about one week from January 12th 1999 to January 14th 1999. The intervals marked by A, B, C and D consist of 2000 ticks and we estimate potential functions directly from these ticks following the method\cite{potentialmodel1}. And in the sub-figures A-1, B-1, C-1 and D-1, we plot $ (P(u)-P_M(u)) $ vs $ (M-1) U(P) $ for each intervals. Here, the super-moving average $P_M$ is defined as $\displaystyle{P_M=\sum_{k=0}^{M-1}P(u-k)}$ where the size parameter  $M$ is 16. $U(P)$ denotes the potential function, and these variables are assumed to satisfy the following equation with a stochastic noise term, $ F(u) $:

 \begin{eqnarray}
P(u+1)=P(u)-\frac{\partial}{\partial P}U(P)\mid_{P=P(u)-P_M(u)}+F(u)\label{eq:PM-1}\\
U(P)=\frac{b(u)}{2(M-1)}(P(u)-P_M(u))^2.
\end{eqnarray}
 
As known from these figures we can approximate the plots by quadric potentials and estimate the curvature of the potential, $b(u)$, as a function of time. 

In the intervals A and B the curvature $b(u)$ is positive and the price fluctuation is attracted by the center of price $P_M$. The diffusion of price fluctuation is slower than a simple random walk case with the same volatility. On the other hand in the intervals C and D, the potential coefficient $b(u)$ is negative and the diffusion is faster. As shown the figures, we can clear time dependent potentials in market prices. It is confirmed that for pure random walk data or randomly shuffled data the estimated potential coefficient distributes around 0 with the standard deviation about 0.2. 

\section{Comparison of potentials with the deterministic dealer model} \label{sec:DM-potential}
In this section we apply the potential analysis to the time series produced by the deterministic dealer model. In this model we assume that this artificial market consists of $N=300$ dealers and every dealer changes his bid price $p_i(s)$ at time $s$ described by the following equation, 
\begin{eqnarray}
\frac{dp_i(s)}{ds}=\sigma_i(s)c_i+d_i<\Delta P>_M.\label{eq.DM-1}
\end{eqnarray}
Here, $\sigma_i$ represents the $i$-th dealer's position,
\begin{eqnarray}
\sigma_i=\left\{
\begin{array}{l}
{+1}\quad \text{buyer}\nonumber\\
{-1}\quad \text{seller}\nonumber
\end{array}
\right. . 
\end{eqnarray}
The parameter $c_i$ is the amount of change at each time step: if $i$-th dealer is a buyer, he monotonically raises his price until he can buy while if he is a seller he reduces his price until he can sell. We set $c_i$  uniform random numbers given initially to represent dealers' character, namely, a dealer who has big $c_i$ is regarded as quick-tempered. The second term $d_i<\Delta P>_M$ represents the foreseeing effect defined in the following. 
\begin{equation}
<\Delta P>_M=\frac{2}{M(M+1)}\sum_{k=1}^{M}(M-(k-1))\Delta P(u-(k-1)),\label{eq:WMA-1}
\end{equation}
where $\Delta P(u-k)=P(u-k)-P(u-(k+1))$ is price change before $k$ ticks. This is a weighted moving average of price changes for past $M$ ticks with bigger weights to newer price changes. For simplicity we set the value of $d_i$ constant for all dealers, that is, $d_i=d$. When $d$ is positive, the dealers are trend followers who predict the near future market price change to be proportional to the latest price change.  When $d$ is negative, the dealers are contrarians who predict in the opposite way, for example, when the market price is going up he predicts that the price will go down in the near future. It is possible to tune the values of parameters so that some of the empirical laws of financial markets are reproduced by this formulation \cite{dealermodel1}\cite{dealermodel2}. 

In this model the $j$-th dealer's ask price is given by $p_j(s)+L$ assuming that the spread is a constant, independent of the dealers. In this artificial market a transaction occurs when a bid price and an ask price match. This condition is described as follows:
 \begin{equation}
 \max\{p_i(s)\}\le \min\{p_j(s)+L\},
 \end{equation}
where $\max\{p_i(s)\}$ is the maximal value of bid prices and $\min\{p_j(s)+L\}$ is the minimal value of ask prices.  Transaction occurs between the pair of dealers who quote the highest bid price and lowest ask price, and the market price $P(u)$ is defined by the middle value, $[\max\{p_i\}+\min\{p_i+L\}]/2$. The dealers are assumed to trade unit volume and after the trade the dealers change their positions, so the buyer becomes a seller and the seller becomes a buyer. In other words, the sign of $\sigma_i$ is reversed when the $i$-th dealer is involved in a transaction. By repeating these deterministic evolution rules this artificial market keeps producing apparently random market prices continuously.

\begin{figure}[htbp]
  \begin{center}
    \includegraphics[width=13.0cm]{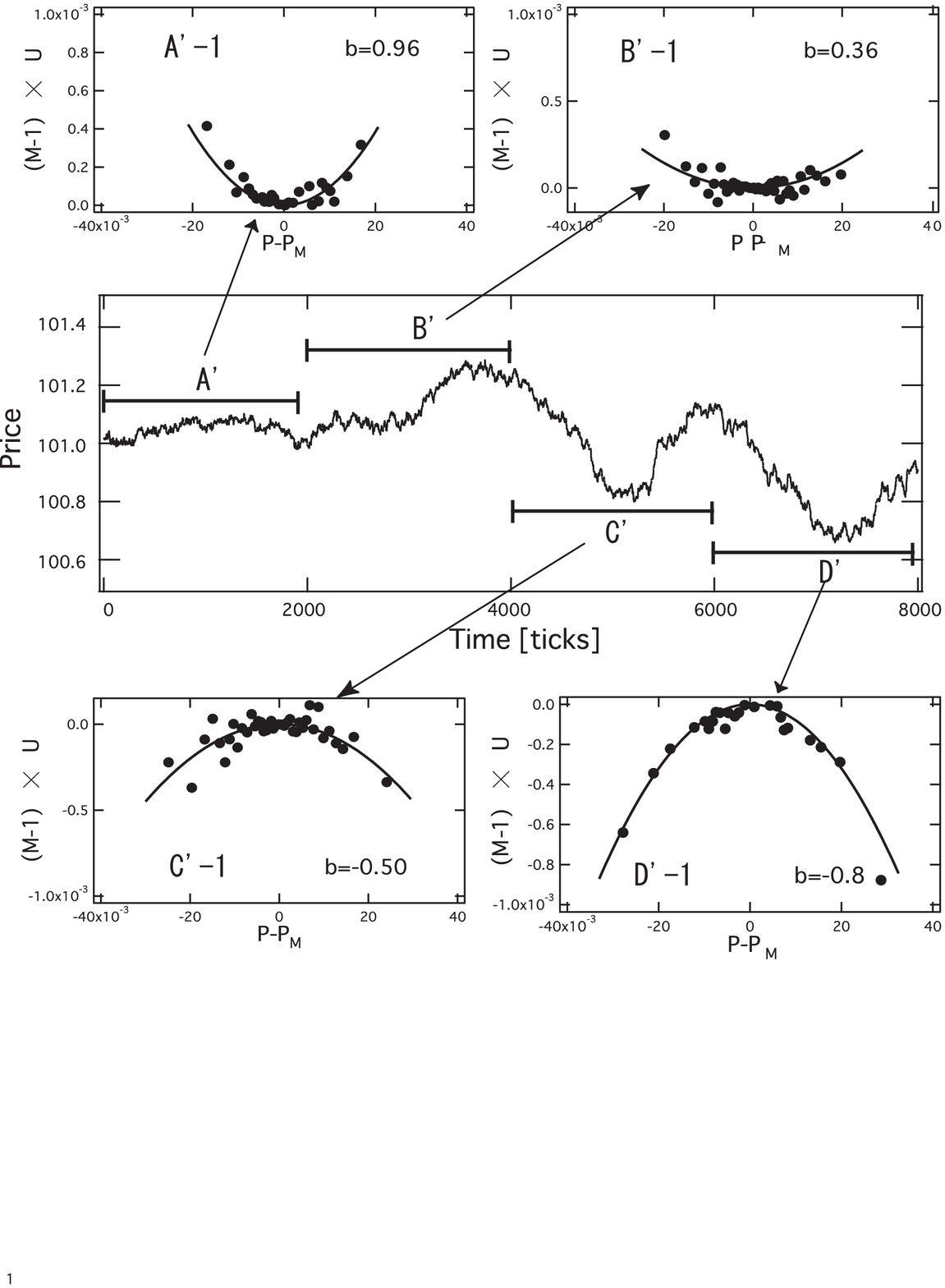}
  \end{center}
  \caption{Reproduced potentials by deterministic dealer model. The ranges marked A$^{\prime}$, B$^{\prime}$, C$^{\prime}$, D$^{\prime}$ are 2000 ticks as well as Fig.\ref{fig:potential_bloom1.eps}.We change the foreseeing parameter $d$ for each range. In range A$^{\prime}$, $d$ is -0.5. In other ranges B$^{\prime}$, C$^{\prime}$, D$^{\prime}$ we change $d$=0, 0.5, 1.0 respectively. For other parameters we identically set as follows: $N$ (the number of dealer)  is 300, $L$ (spread) is 1.0 and the range of $c_i$ is [0.01,0.02].}
  \label{fig:DM3-potential1.eps}
\end{figure}

We apply the potential analysis method previously described to the resulting artificial market price data. For any value of the foreseeing parameter $d$ we can observe non-trivial potential forces and the value of potential coefficient can be estimated. Here, we show four typical cases marked A$^{\prime}$, B$^{\prime}$, C$^{\prime}$, D$^{\prime}$ in Fig.\ref{fig:DM3-potential1.eps}.  In these simulations all parameters except $d$ are identical. In the interval of Fig.\ref{fig:DM3-potential1.eps} A$^{\prime}$, the parameter $d$ is negative, that is, dealers are contrarians and we find a stable potential as shown in Fig.\ref{fig:DM3-potential1.eps} A$^{\prime}$-1. When the parameter $d$ equals zero, the observed potential function is nearly flat as shown in Fig.\ref{fig:DM3-potential1.eps} C$^{\prime}$-1. When the parameter $d$ is greater than zero, the dealers are trend-followers and we observe an unstable potential as shown in Fig.\ref{fig:DM3-potential1.eps} C$^{\prime}$-1 and D$^{\prime}$-1. From these results we understand that the trend-following effect directly contributes to the unstable potential and the contrarian effect is responsible for the stable potential. 

\begin{figure}[htbp]
  \begin{center}
    \includegraphics[width=8.0cm]{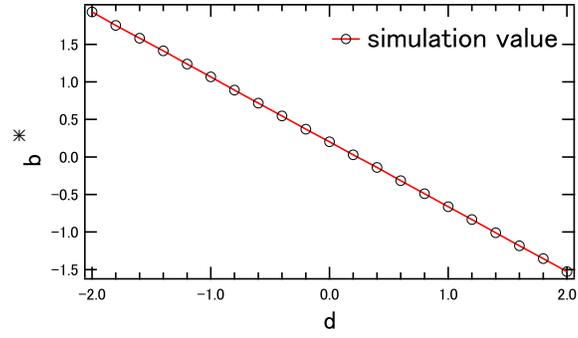}
  \end{center}
  \caption{The relation between potential curvature $b^{\ast}$ plotted horizontal axis and dealers' foreseeing effect $d$ plotted bottom line. The $b^{\ast}$ means the average of tick time for $b$. In this simulation we set the parameters as follows: $N$ (the number of dealer)  is 300, $L$ (spread) is 1.0 and the range of $c_i$ is [0.01,0.02]. }
  \label{fig:relation-b-d.eps}
\end{figure}

In Fig.\ref{fig:relation-b-d.eps} we plot the estimated value of the potential coefficient $b^{\ast}$ taken average over $b(u)$ for 100,000 ticks as a function of the foreseeing parameter $d$. All the points are clearly on a straight line $b^{\ast}=0.2-0.86d$. 
When $d=0$, we can observe very weak attractive potential force, $b^{\ast}=0.2$ by taking average over long numerical simulation. We conjecture that this attractive force is not a numerical error but it is caused by the transaction rule. In our model a buyer becomes a seller and a seller becomes a buyer after the transaction. Due to this effect of changing roles between two dealers, it pulls back the market price to the mean values of the past prices. It can be described by the random walk in the attractive force.

\section{conclusion and discussion}
In this paper we showed by numerical simulations that dealers' foreseeing effect is the source of the market potential force observed in real financial markets. As a result, we have found that the coefficient of the potential function and the dealers' foreseeing action is deeply related. When the dealers are trend-followers the potential coefficient is negative and the market is unstable, and when the dealers are contrarians the coefficient is positive and the market is stable. In real markets the coefficient value is always fluctuating as typically shown in Fig.1. This implies that the prediction strategy of dealers in real markets may not be fixed to either trend-followers or contrarians, but they change the strategy from time to time. 

For real market dealers it is said that to know other dealers' strategy or the whole market atmosphere is very important. It may become possible to tell by using the coefficient $b$ whether the market is dominated by trend-followers or contrarians in real time.

\begin{acknowledgements}
\end{acknowledgements}


\end{document}